\documentclass[sigconf]{acmart}
\AtBeginDocument{%
  }

\copyrightyear{2026}
\acmYear{2026}
\setcopyright{cc}
\setcctype{by}
\acmConference[ACI 2026]{ACI ’26: International Conference on Animal-Computer Interaction}{December 02--05, 2026}{Windisch, Switzerland}
\acmBooktitle{ACI ’26: International Conference on Animal-Computer Interaction (ACI 2026), December 02--05, 2026, Windisch, Switzerland}
\acmDOI{10.1145/3859975.3859990}
\acmISBN{979-8-4007-2558-6/2026/12}

\usepackage{subcaption}
\begin{document}

\title[Robot Navigation and Canine Welfare]{Robot Navigation and Canine Welfare: A Pilot Study Surfacing Challenges for Cross-Species Legibility and Proxy Validity in Animal–Machine Interaction}

\author{Haoxiang Zhang}
\email{hzhang68@pratt.edu}
\orcid{0009-0001-0556-1152}
\affiliation{%
  \institution{Pratt Institute}
  \city{New York}
  \state{New York}
  \country{USA}
}

\author{Nancy Smith}
\affiliation{%
  \institution{Pratt Institute}
  \city{New York}
  \state{New York}
  \country{USA}
}
\email{nsmith7@pratt.edu}
\orcid{0000-0002-2017-2921}

\renewcommand{\shortauthors}{Zhang and Smith}

\begin{abstract}
  Autonomous delivery robots increasingly share residential spaces with companion animals, yet navigation behavior is rarely designed with animal welfare in mind. This paper presents a pilot study evaluating a curve-out maneuver, a laterally arcing pass-by trajectory triggered at the dog's alert distance, proposed as a candidate welfare-informed navigation intervention grounded in canine social communication research and legible motion theory. Following a research through design approach, we developed and evaluated a first design iteration of the artifact in a naturalistic home environment with a companion dog. Contrary to our hypothesis, the curve-out produced a more acute stress response than the straight-line baseline. We argue that discontinuous execution limitation, the robot stopping and reorienting multiple times to approximate the arc, undermined legibility, which in itself became a stressor. The pilot surfaces the legibility of robot behavior to non-humans’ eyes and proxy stimulus validity as structural challenges for AMI and repositions the design problem for future work.
\end{abstract}

\begin{teaserfigure}
  \centering
  \includegraphics[width=0.6\textwidth]{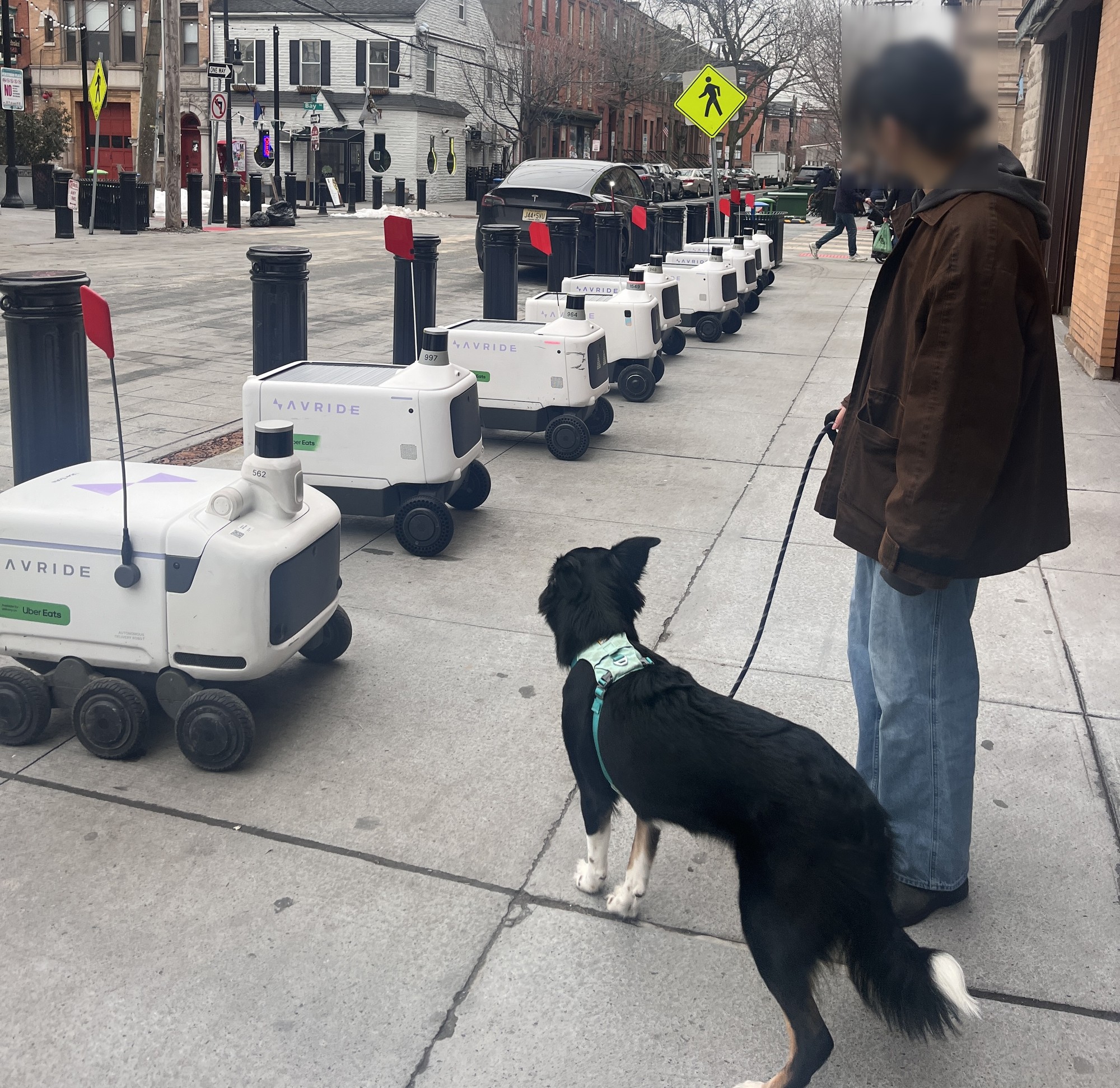}
  \caption{A companion dog encountering a fleet of autonomous delivery robots parked on a residential sidewalk.}
  \Description{A companion dog on a residential sidewalk facing a group of parked autonomous delivery robots.}
  \label{fig:teaser}
\end{teaserfigure}

\maketitle

\section{Introduction}
Autonomous delivery robots, designed to deliver food and groceries, are increasingly being deployed in residential neighborhoods \cite{lakshmi2020understanding}. As these systems move beyond controlled industrial contexts into urban spaces of daily living, they encounter not only humans but also companion animals. Väätäjä et al. \cite{vaataja2023exploring} report that about 17\% of the dog owners surveyed found their dog reacted negatively when first encountering autonomous delivery robots, some developing persistent adverse associations. Those unexpected encounters signal the situation where the technology was not designed for animals but affects them anyway \cite{10.1145/3637882.3637895}, positioning dogs as usees who are affected by those robots and have no agency to opt out, negotiate, or influence the robot's behavior \cite{hirskyj2021forming}. Animals in general appear to be excluded from obstacle-detection design: commercial delivery robot operators describe their sensors as targeting people, cars, and obstacles~\cite{uber2026robots}, seemingly classing companion animals alongside inert obstacles rather than as agents the system is designed to perceive and respond to. As a result, the burden of adapting to unpredictable robot behavior falls entirely on the animals. This work takes an early step toward redistributing that burden to the machine through deliberate design, using a single navigation intervention as a probe to surface the challenges such redistribution entails.

Following a research through design approach \cite{10.1145/1240624.1240704} \cite{10.1145/2556288.2557357}, this paper proposes an animal-welfare-informed robot navigation behavior, a curve-out maneuver, as a candidate design intervention for autonomous robots operating in animal-inhabited urban environments. This design is informed by ethological research on canine social communication, which identifies curved and indirect approaches as stress and conflict-reducing signals in multi-species interaction \cite{siniscalchi2018communication} \cite{FIRNKES201735}, and in the legible motion theory of human-robot interaction \cite{10.5555/2447556.2447672}, which shows that the behavior design could communicate the intention of the agent to an observer. We treat trajectory geometry as a spatial design parameter and the curve-out as an artifact intended to transform a current problematic state, which is identified as a stress-triggering direct passing-by robot navigation behavior, to a preferred one, where robot movement actively signals non-threat and supports coexistence in a compact urban environment. The artifact represents a first design iteration, intended to be refined through subsequent cycles informed by pilot observation. A pilot evaluation was conducted in a naturalistic home environment with a companion dog with documented adverse reactions to delivery robots. 

By isolating trajectory geometry as a spatial design parameter and conducting a structured pilot evaluation, this work contributes to the emerging Animal Machine Interaction agenda, surfaces cross-species legibility design as a candidate mechanism for future investigation, and documents the methodological challenges of evaluating welfare-informed robot behaviors with companion animals.

\section{Background}
The existing research body has examined interactions between dogs and robotic systems primarily in controlled experimental settings. Studies have used social robots such as AIBO \cite{kubinyi2004social} and embodied robotic agents \cite{qin2020dog}, as well as remote-controlled toy cars and other unfamiliar moving objects \cite{gergely2013you} \cite{GERGELY2015137}, to assess behavioral reactivity, emotional responses, and social engagement. Research on dogs with elevated anxiety has further shown that robot movement can function as a genuine stress stimulus, producing measurable behavioral inhibition, including reduced movement and avoidance in affected individuals \cite{zamansky2018effects}. Moreover, Torjussen et al. \cite{10.1145/3637882.3637896} found that even dogs with no prior history of separation-related distress exhibited measurable increases in distress behaviors when exposed to stimuli from remote pet-monitoring devices compared to a no-stimulus control. However, these controlled contexts differ substantially from real-world encounters, where companion dogs meet autonomous robots incidentally. 

Väätäjä et al. \cite{vaataja2023exploring} conducted an online survey with dog owners whose dogs had encountered delivery robots during a grocery delivery pilot, collecting responses for 212 dogs. They found that about 17\% responded negatively at their first encounter, including intensive barking and retreating, with some developing persistent adverse associations, forcing some owners to avoid certain narrow sidewalks where those robots frequently appear. The authors proposed that the robots' straight-line trajectories and sudden changes in movement may have contributed to these stress responses, and that individual temperament may have accounted for some of the variance in reactions. Since the study relied on owner self-reports, however, these explanations remained hypothetical, and no controlled study had been conducted to test them.

The trajectory explanation proposed by Väätäjä et al., \cite{vaataja2023exploring} could draw theoretical support from canine social communication research. In dog-dog and human-dog interaction, straight-line approaches combined with sustained frontal orientation have been associated with assertiveness or threat signaling, whereas curved or indirect approaches function as tension-reducing behaviors \cite{siniscalchi2018communication} \cite{FIRNKES201735} \cite{ricci2016effects}. This body of work provides a principled basis for hypothesizing that trajectory geometry may explain how dogs respond to approaching robotic agents. Whether dogs apply the same social interpretive framework to machines as they do to other living beings, however, remains an open question. Gergely et al.'s \cite{gergely2013you}, \cite{GERGELY2015137} findings suggest that dogs responded differently to self-propelled objects depending on whether those objects displayed goal-directed or contingent behavior, indicating that dogs actively attempt to classify agents based on movement qualities, though the interpretive framework they apply to robotic agents remains unresolved. The canine social communication literature, therefore, motivates trajectory geometry as a possible candidate design parameter that worth exploring.

We found that comparatively little work has examined how autonomous machines should navigate in animal-inhabited urban environments to protect animal welfare and reduce stress. To our knowledge, trajectory geometry has not been experimentally isolated as an independent design parameter in canine–robot interaction with companion dogs.

From an AMI perspective, trajectory design is a matter of coexistence in shared compact urban environments and animal-friendly system design. If certain approach geometries resemble assertive or threatening signals in canine perception, then navigation behavior becomes an ethical and welfare-relevant design concern. This work, therefore, treats passby trajectory geometry as a welfare-relevant design parameter and proposes the curve-out maneuver as a concrete design response, an artifact grounded in both canine social communication research and legible motion theory, intended to make robot navigation behavior more animal-friendly in shared residential environments.

\section{Design and evaluation approach}

This section documents the design of the curve-out navigation artifact, the rationale for each design decision, and the structured pilot observation approach used to evaluate the artifact in context.

\subsection{Participant}
The pilot evaluation was conducted with an 11-month-old healthy female shepherd mix living in the researcher's residential home environment. The participant has maintained a mild-moderate adverse response to neighborhood delivery robots, including freezing, barking, and retreating, despite positive counter-conditioning. The owner completed the Canine Behavioral Assessment and Research Questionnaire (C-BARQ), which indicated elevated scores on the stranger-directed fear and non-social fear subscales relative to breed averages. This dispositional profile is consistent with Väätäjä et al., \cite{vaataja2023exploring}'s speculation that individual temperament contributes to variance in dog-robot encounter outcomes, and makes her a well-characterized evaluation case for this design contribution. Consistent with a research through design approach, the pilot observation was conducted with a single participant; the goal is to assess the artifact's real-world behavior and generate design implications, not to produce generalizable empirical claims.

Prior to pilot evaluation, the participant was habituated to the stationary platform through repeated non-contingent exposure in the home environment. This is to ensure that the observed responses reflected movement behavior rather than residual novelty responses to the object itself.

\subsection{Evaluation Setup and Trajectory Design}
\begin{figure}[htbp]
  \centering
  \includegraphics[width=\linewidth]{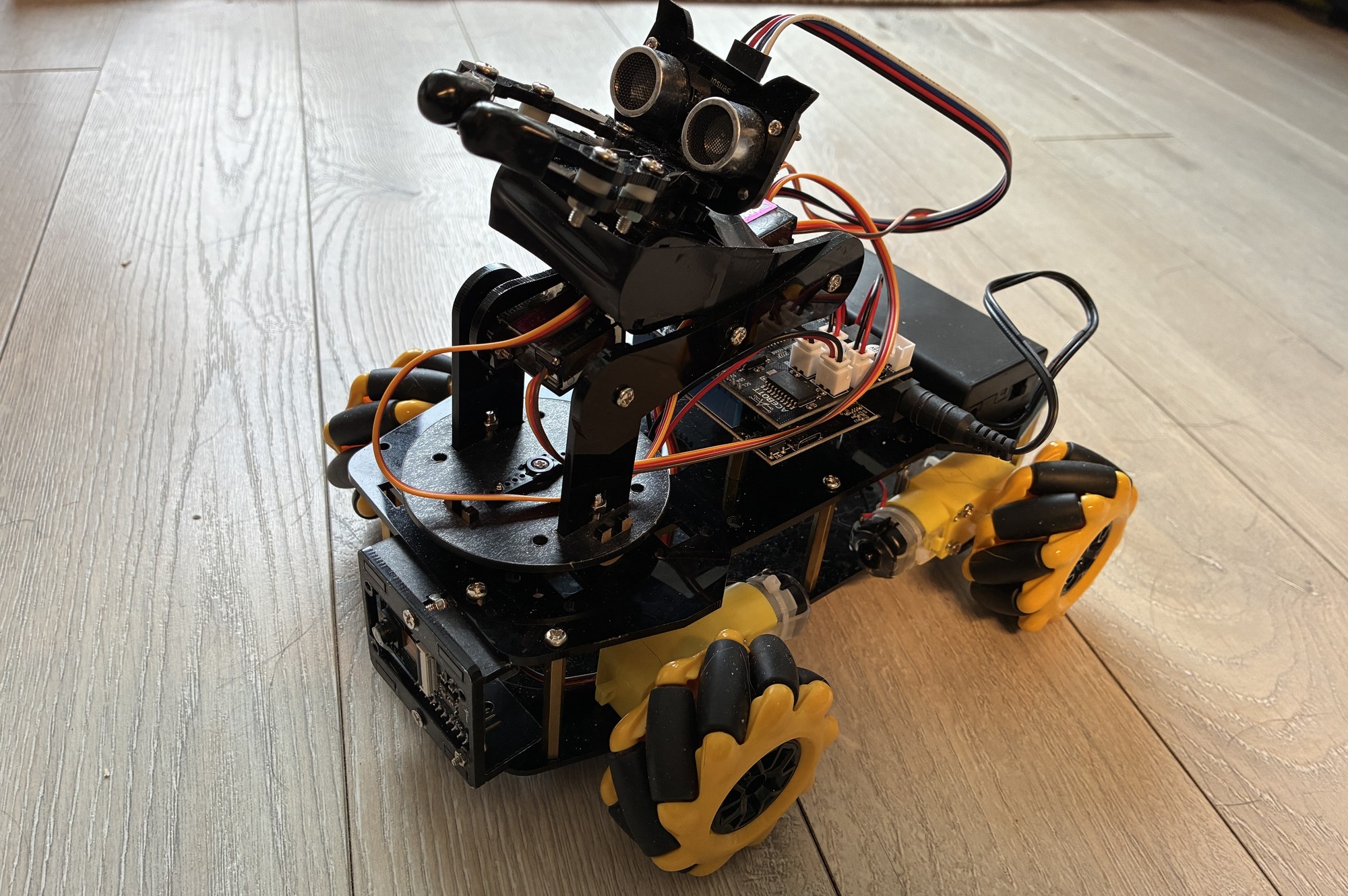}
  \caption{The ESP32-based mobile platform used as the robotic proxy stimulus. The arm attachment was not used in this evaluation.}
  \Description{A motor-based ESP32 platform used as the robotic proxy stimulus.}
  \label{fig:esp32}
\end{figure}

The artifact is implemented using an ESP32-based mobile platform (Figure~\ref{fig:esp32}). The platform is a commercially available robotics chassis, modified with additional off-the-shelf components. It measures 25 × 18 × 18 cm (L × W × H), with a black chassis and yellow mecanum wheels. The platform traveled at a fixed speed of approximately 30 cm/s (motor input value 180). Given the constraints of the control library used in this pilot, which only supports sequenced heading/duration commands, the platform can only execute movement patterns in stop-turn-move segments, a limitation discussed more closely in sections 5 and 6. The platform produces an audible electric motor whir and wheel-surface contact noise (mecanum wheels on hardwood flooring). It does not produce distinct alert tones or beeps. 

The artifact's behavior execution is split across two layers. A Python-based controller, running on a connected laptop, automates the computation the segment geometry for a given trial (turning angles, chord lengths, and segment durations) from runtime inputs, including the estimated alert distance and the desired segment count, using the geometry formula described below (Equation~\ref{eq:chord} and Table~\ref{tab:turning-angles}). This controller then transmits a sequence of discrete motor commands (heading and duration) to the ESP32 platform via ESP-NOW, using the chassis's included control library largely unmodified; the platform itself performs no path planning and simply executes each command in sequence. Speed and rotation timing (30 cm/s at motor input 180, and 1000 ms per 90 degree pivot at the same motor input) were established through a separate calibration procedure prior to the pilot, and are used by the controller to convert desired distances and angles into motor command durations.

The pilot evaluation sessions took place in the participant's primary residential living space. It was a decision made for two reasons. First, a familiar environment minimizes confounding environmental variables, such as ambient noise, unfamiliar smells, other pedestrians or animals that would be uncontrolled in a public sidewalk setting and could independently elevate arousal or introduce variance unrelated to the trajectory manipulation itself. Second, given the participant's documented anxious temperament (Section 3.1) and the repeated-exposure design required by the alternating trial structure, conducting the pilot in an unfamiliar or high-stimulus public environment risked compounding stress from novelty and lack of escape options with the stress under investigation, raising a welfare concern independent of the study's validity. The home environment, paired with the familiar dog bed as a safety retreat zone (Section 3.2), allowed repeated trials while keeping the participant's overall stress exposure within what her established temperament profile could reasonably tolerate. 

\begin{figure}[htbp]
  \centering
  \begin{subfigure}{0.45\linewidth}
    \includegraphics[width=\linewidth]{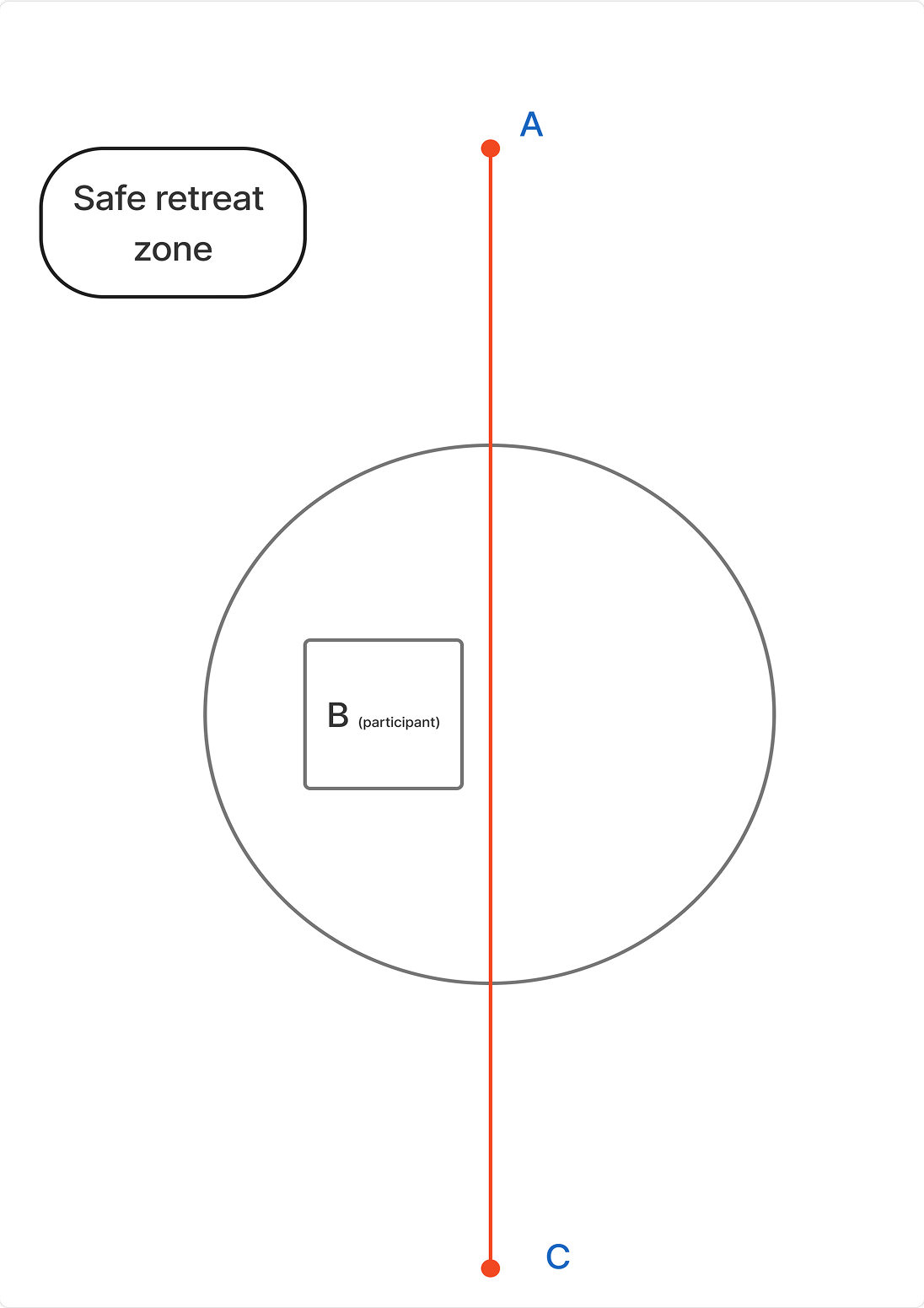}
    \caption{Baseline trajectory}
    \label{fig:baseline}
  \end{subfigure}
  \hfill
  \begin{subfigure}{0.45\linewidth}
    \includegraphics[width=\linewidth]{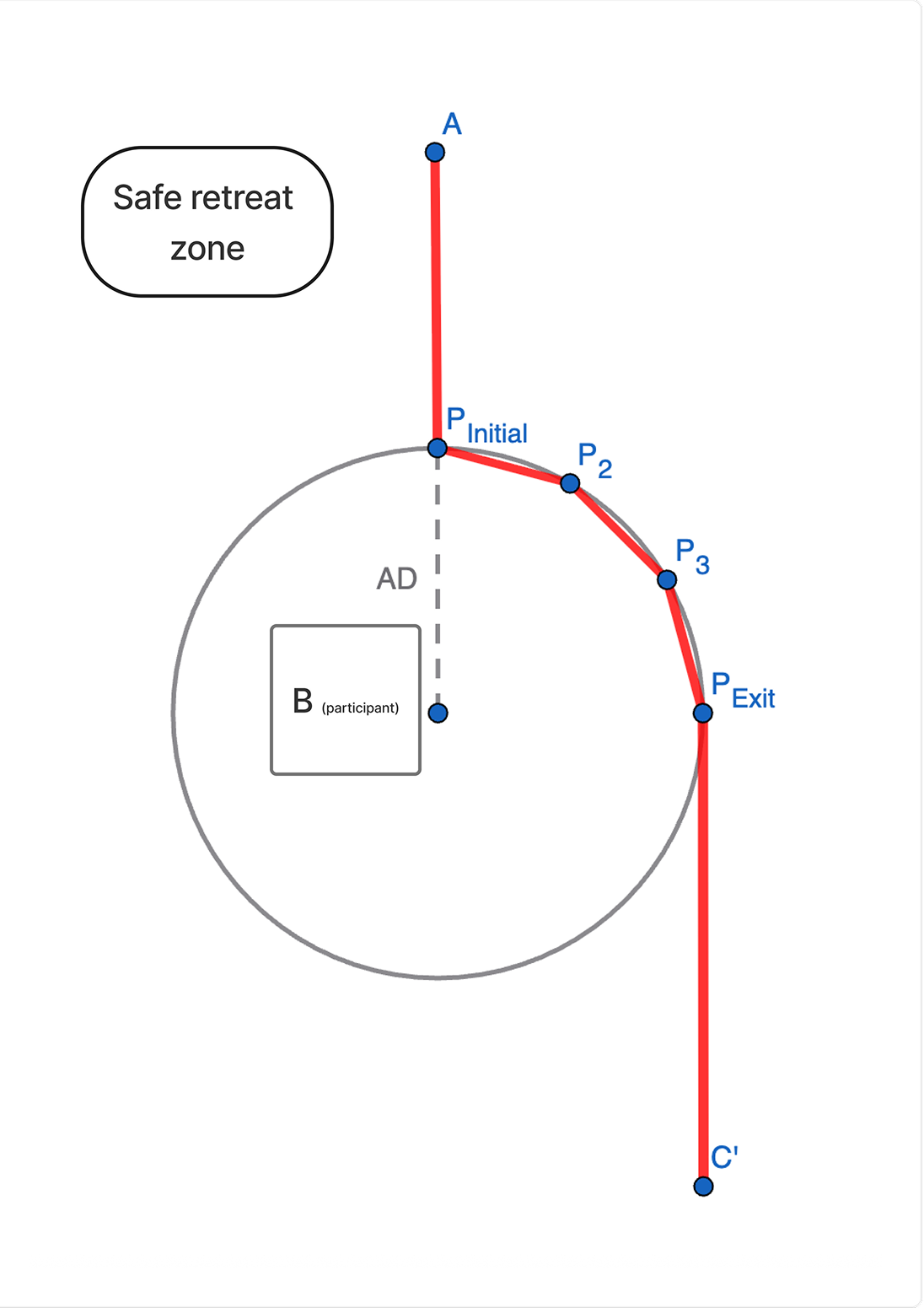}
    \caption{Curve-out trajectory}
    \label{fig:curveout}
  \end{subfigure}
  \caption{Designed trajectories: (a) baseline straight-line pass-by and 
  (b) curve-out maneuver triggered at alert distance.}
  \Description{Two trajectory diagrams. Left: the robot follows a straight 
  line from A through participant position B to endpoint C. Right: the robot 
  follows a straight line from A, then curves outward at the alert distance, 
  arcing away from B before continuing to endpoint C prime.}
  \label{fig:trajectories}
\end{figure}

The setup remained consistent across all sessions. The robot travels from point A to point C, passing the participant's fixed position at point B. A training mat marked the participant's designated starting position at point B, and a dog bed was placed outside of the testing environment as a safety retreat zone, providing a familiar refuge should she choose to disengage at anytime (Figure~\ref{fig:baseline})

\textbf{Alert Distance estimation}
Prior to pilot evaluation sessions, a preliminary session was conducted to estimate the participant's alert distance (AD): the radial threshold at which the participants notice the stimulus and start exhibiting an observable stress signal, such as sustained orientation, postural stiffening, or retreat initiation. This threshold informed the timing of the curve-out maneuver in the designed trajectory and established a consistent perceptual reference point across sessions. 

\textbf{Baseline trajectory}
In the baseline trajectory, the robot followed a fixed linear path from start to end without deviation, passing the participant at the closest point of approach (Figure~\ref{fig:baseline}). This trajectory represents the direct pass-by geometry of current autonomous delivery robots. 

\textbf{Curve-out trajectory}
In the curve-out trajectory, the robot followed the same linear path until reaching the estimated AD, at which point a hard-coded curve-out maneuver would be triggered. The robot would then arc laterally away from the participant's position and continue to the endpoint C’ on the adjusted path without returning to the original linear trajectory (Figure~\ref{fig:curveout}). The curve-out geometry is designed to approximate the indirect, tension-reducing approach trajectories observed in canine social interaction, in which curved or averted movement functions as an appeasement signal \cite{siniscalchi2018communication} \cite{ricci2016effects}.

By triggering the maneuver at AD, the point at which the participant first detects and responds to the approaching object, the design aims to produce legible and stress-reducing behavior right before the onset of the stress response.

Table~\ref{tab:turning-angles} summarizes the turning angle for each phase. 
The chord length $\ell$ traveled by the robot per segment is determined by the alert distance and the number of slices:
\begin{equation}
  \ell = 2 \cdot AD \cdot \sin\left(\frac{45^\circ}{N-1}\right)
  \label{eq:chord}
\end{equation}
where $AD$ is the estimated alert distance, $N$ is the number of discrete 
segments, and $\ell$ is the chord length traveled per segment.

\begin{table}[h]
  \centering
  \caption{Turning angles for the curve-out maneuver ($N$ = number of stops the robot takes to turn)}
  \label{tab:turning-angles}
  \renewcommand{\arraystretch}{1.4}
  \begin{tabular}{ll}
    \toprule
    \textbf{Phase} & \textbf{Angle (Degree)} \\
    \midrule
    Initial turn & $90^\circ - \frac{45^\circ}{N-1}$ \\
    Middle steps & $\frac{90^\circ}{N-1}$ \\
    Exit turn & $\frac{45^\circ}{N-1}$ \\
    \bottomrule
  \end{tabular}
\end{table}

In the pilot implementation, $AD = 60$ cm and $N = 4$ steps were used, yielding an initial right turn of $75^\circ$, two middle segments of $30^\circ$ each, and a final left turn of $15^\circ$, with a chord length of $\ell = 2 \times 60 \cdot \sin(15^\circ) \approx 31$ cm per segment. The choice of N = 4 was informed by an empirically observed tradeoff: a higher segment count would more closely approximate a continuous arc but require more discrete stop-rotate-move transitions, while a lower count would reduce transitions at the cost of a coarser approximation. 

\subsection{Evaluation procedure}

Once the AD was established, the evaluation session followed the same protocol. It began with the participant being brought into the evaluation space and allowed to settle at the designated training mat at point B. The training mat serves as a spatial anchor for the participant's starting position, providing a consistent orientation relative to the robot's path across sessions. Trials were administered in an alternating sequence (baseline trajectory, curve-out trajectory, baseline trajectory, curve-out trajectory) to allow structured comparison between current robot navigation behavior and the designed intervention across repeated observations. The starting trajectory was determined by a coin flip prior to the first session. Following each trial, a mandatory inter-trial interval was observed until the participant met the behavioral reset criterion described below. Sessions would be terminated according to the stopping criteria defined below, or when a minimum of three trials per trajectory have been completed across sessions. 

\textbf{Between-trial reset}
A mandatory inter-trial interval was observed immediately after the robot reached endpoint C or C’ and lasted until the participant returned to a defined behavioral baseline: four paws on the ground, no sustained orientation toward the robot, and the capacity to accept a food treat from the handler. Trials only proceeded until this criterion was met. This reset procedure ensures that observations in successive trials reflect the designed trajectory rather than carry-over arousal from the preceding trial.

\textbf{Session stopping criteria}
Individual trials would be terminated immediately if the participant exhibits an intense stress response, including sustained retreat to a fixed location with refusal to reorient, or continuous vocalization lasting more than 10 seconds. Sessions would be terminated if the stopping criteria were triggered or if the participant was unable to meet the between-trial reset criterion within 10 minutes. 

\subsection{Observational Documentation}
Canine behavioral responses were coded in real time by the observer using an ethogram. The onset and offset of key behaviors were logged during each trial after the robot passed AD. Two behavioral indicators were documented to assess how the artifact performs in the context.

The first indicator was stress-related behavior. Coded behaviors included retreat, which was operationally defined as slowly moving away from the robot; sustained orientation, which was defined as fixed orientation toward the robot while the participant could still shift weight, reposition, or pant normally; freezing, which defined as the state when the participant pauses movement entirely, lower her body, and reduced or suspended panting; vocalization, which includes barking and whining; and lastly, escape, which was defined as a high-speed, intense movement moving away from the robot. Behaviors were documented from onset to offset, and co-occurring behaviors were coded simultaneously. A reduction in these behaviors during curve-out passes relative to straight-pass trials would indicate the artifact is moving toward the preferred state.

The second indicator was post-trial recovery, documented as the interval between the robot's exit from the participant's visual field and her first successful acceptance of a food treat from the handler. Treat acceptance: taking and consuming the treat without redirecting attention to the robot is a well-validated behavioral marker of reduced arousal in dogs.

\subsection{Ethical considerations}
This evaluation was designed in accordance with animal welfare principles. The participant was exposed only to a self-propelled mobile platform in her familiar home environment, a stimulus analogous to objects she encounters in routine daily life. The participant's behavioral responses function as ongoing indicators of willingness to continue participation. Specifically, the stopping criteria serve as her primary exit mechanism: trials would be terminated immediately upon signs of acute distress, and sessions would be terminated if distress criteria are triggered on 2 consecutive trials. The dog bed provides a physical refuge she can access at any time, further operationalizing her agency to disengage. 

\section{Pilot Observation}

\subsection{Alert Distance Estimation}
Three AD estimation sessions were conducted within a single day. In the first session, the participant stood up and retreated when the stimulus was approximately 96 cm away, recovering almost immediately; as this constituted her first exposure to the stimulus in motion, the response was treated as a novelty artifact. On re-exposure, the participant stood up and fixated on the stimulus at approximately 60 cm, showing no escalation beyond alert behavior. A third session confirmed this threshold. The operative AD for all subsequent trials was set at 60 cm.

\subsection{Baseline Trajectory}
Each baseline trial spanned a fixed distance of 400 cm at the platform's calibrated travel speed of 30 cm/s, corresponding to a trial duration of approximately 13.3 seconds.

Three baseline trials were conducted. In the first baseline trial (trial 1), the participant showed sustained orientation ($\sim$1s) and freezing ($\sim$1s) toward the platform, escalating to retreat ($\sim$3s) directed away from it; ear pricking, a notable active postural effort in this floppy-eared individual, co-occurred with freezing. Recovery to treat acceptance took approximately 30 seconds. Once the platform had passed and increased its distance, the participant turned and approached it from the rear, suggesting her response might've combined avoidance with curiosity rather than sustained fear.

In the second baseline trial (trial 3), the participant showed sustained orientation ($\sim$5s) without escalation to freezing or retreat, and recovered within approximately 12 seconds. By the third baseline trial (trial 5), orientation was brief ($\sim$1s) and recovery was immediate, consistent with habituation within a single session, faster than the alternating trial design had anticipated.

No vocalization was observed across 3 baseline trials. 

\subsection{Curve-out Trajectory}
Using the empirically established AD of 60 cm (Section 4.1) and N = 4 segments, each curve-out trial consisted of a straight approach to the alert-distance threshold, three chorded arc segments separated by in-place pivots (833 ms, 333 ms × 2, 167 ms), and a straight exit leg, for a total nominal duration of approximately 15.6 seconds

Three curve-out trials were conducted. In the first curve-out trial (trial 2), the participant's response escalated further than in any baseline trial: sustained orientation ($\sim$2s) and freezing ($\sim$3s) preceded escape ($\sim$5s), and recovery took approximately 37 seconds, the longest of the six trials. In the second curve-out trial (trial 4), orientation ($\sim$4s) and freezing ($\sim$2s) again preceded a flight response, though of lower intensity (retreat), with recovery of approximately 22 seconds. Notably, escalation to a terminal flight behavior took longer to develop in both curve-out trials than in the first baseline trial, despite reaching a more intense terminal behavior in trial 2 (escape) than baseline ever reached (retreat and approach); this pattern is discussed further in Section 5. No voluntary investigative behavior was observed following either curve-out trial. By the third curve-out trial (trial 6), orientation was brief ($<$1s) and recovery was immediate, indicating habituation to the curve-out trajectory occurred one trial later than for the baseline condition, closing the observation window before additional comparative trials could be recorded.

No vocalization was observed across 3 curve-out trials. 

\begin{table*}[t]
\caption{Coded behavioral sequences across trials, ordered chronologically. Sequences show behaviors in the order observed, with approximate durations; ``$\rightarrow$'' indicates one behavior followed another, ``+'' indicates behaviors co-occurred. (Section 3.4).}
\label{tab:behavior-sequences}
\resizebox{\textwidth}{!}{%
\begin{tabular}{|c|l|l|c|}
\hline
\textbf{Trial \#} & \textbf{Condition} & \textbf{Behavior Sequence} & \textbf{Recovery to Treat } \\
\hline
1 & Baseline & Sustained orientation ($\sim$1s) + Freezing ($\sim$1s) $\rightarrow$ Retreat ($\sim$3s) $\rightarrow$ Approach from rear ($\sim$5s) & $\sim$30s \\
\hline
2 & Curve-out & Sustained orientation ($\sim$2s) + Freezing ($\sim$3s) $\rightarrow$ Escape ($\sim$5s) & $\sim$37s \\
\hline
3 & Baseline & Sustained orientation ($\sim$5s), no Retreat or Escaping & $\sim$12s \\
\hline
4 & Curve-out & Sustained orientation ($\sim$4s) + Freezing ($\sim$2s) $\rightarrow$ Retreat ($\sim$2s) & $\sim$22s \\
\hline
5 & Baseline & Sustained orientation ($\sim$1s) & Immediately \\
\hline
6 & Curve-out & Sustained orientation ($<$1s) & Immediately \\
\hline
\end{tabular}%
}
\end{table*}

\section{Discussion and Design Implications}
The pilot observations found that the curve-out trajectory, when executed as segmented approximations, elicited a more acute stress response than the straight-line trajectory, partially contradicting the original hypothesis that the curve-out trajectory would reduce stress in dogs. This observation raises an important caveat regarding whether the manner in which a behavior is executed influences animals’ perception of it. In this case, the segmented curve-out behavior may not have been perceived as a continuous and intentional trajectory, but rather as a sequence of abrupt or unpredictable movements, which therefore became the stressor itself. 

In the context of Human Robot Interaction (HRI), an adjacent field, Dragan et al. \cite{10.5555/2447556.2447672} drew a distinction between predictability and legibility of robot behavior. In their formalism, predictability is a goal-to-trajectory inference: an observer who knows the agent's goal expects a certain path and is unsurprised when the agent takes it. Legibility is the inverse, a trajectory-to-goal inference: the observer watches the path unfold and infers the agent's intent from early movement evidence. The present results, although within the context of canine-machine interaction, show a similar pattern. The straight pass became predictable through repetition: once the participant had established a model of the robot's goal, the direct path matched expectation, and habituation followed quickly. The curve-out was designed to be legible, providing early evidence of non-threat intent, but its execution as a series of discrete stop-rotate-move segments made it neither legible nor predictable. The sudden movement within segments did not allow a reliable goal inference, and it did not match any prior expectation. It failed on both axes simultaneously, and the elevated stress response is consistent with that failure.

This is consistent with at least one owner account in Väätäjä et al.,\cite{vaataja2023exploring} in which a dog that had approached a stationary robot to investigate became frightened and aggressive when the robot abruptly resumed movement. The pattern, an investigative approach interrupted by unpredictable motion followed by persistent aversion, mirrors the sequence observed in the pilot observation. Dragan et al. \cite{10.5555/2447556.2447672} found a related effect with a complex robot: when motion deviated even slightly from observer expectation, confidence in goal prediction dropped sharply, a consequence of a mismatch between the observer's cost function and the robot's actual movement profile. To the extent that a comparable inferential process operates in dogs, the participant in the present study had no established expectation model for the curve-out maneuver, so the segmented arc may have produced exactly this kind of mismatch before any legible signal could be read.

A further complication is that breaking the curve-out's continuous arc into four discrete segments necessarily requires additional time for each segment's start-stop-pivot sequence, resulting in a longer total trial duration than the baseline (approximately 15.6s vs. 13.3s, Section 4.2–4.3). This means the curve-out afforded the participant more total time in the robot's presence than the continuously moving baseline. As shown in the results, it took longer (approximately 5s vs. 2s, Trial 2 vs. Trial 1) for the participant to escalate from stationary reaction (orientation and freezing) to flight state in the first curve-out trial (trial 2), but she ended up in a more intense terminal behavior (escape) than in the first baseline trial (trial 1: retreat and approach). This raises a question for future study to explore: why did the segmented curve-out elicit a slower but stronger response? One possible explanation that potentially aligns with Dragan et al. \cite{10.5555/2447556.2447672}'s framework is that the repeated sequence of stop-turn-move cumulated into greater uncertainty, leading to a more intense terminal behavior. Another possibility is that the longer total duration alone, independent of any particular structural feature of the curve-out, increased the likelihood of escalation simply by extending the time the participant spent in the robot's presence. 

However, whether Dragan et al. \cite{10.5555/2447556.2447672}'s robot legibility framework applies to nonhuman participants remains underexplored. The results of this pilot suggest that the participant's perception of the proxy was not a lower-fidelity version of the human case, but a distinct failure mode this framework was never built to predict. As autonomous systems move into shared urban and domestic spaces, the agents observing and interpreting their motion increasingly include companion animals, wildlife, and livestock, not just people.  Future research could examine whether goal-inference processes comparable to those \cite{10.5555/2447556.2447672} identified in human observers operate in dogs, and under what conditions movement patterns function as legible signals across species. 

Bobu et al.’s \cite{10.1145/3610977.3634987} framework on representation alignment may offer a different perspective. They formalize a representation as the robot’s set of tasks‑relevant features that must be aligned with stakeholders’ underlying objectives so that its behavior is value-aligned and legible during execution. In this pilot, the representation of appeasement signals was approximated from studies on dog–dog and dog–human interaction, but whether dogs hold a similar representation for a robot’s motion is unexplored. Secondly, within the designed representation of the artifacts, motion continuity, which may be a salient task‑relevant feature from the dog’s perspective, was missing from execution. Unfortunately, trajectory smoothing was not evaluated within this pilot. The segment count was decided and executed in the first curve-out trial. Introducing a smoothed or differently-segmented trajectory partway through the pilot would have meant the curve-out trials no longer represented a single, consistent condition, undermining our ability to draw any conclusion about the curve-out maneuver as tested. By the close of this pilot, the participant had habituated to the proxy, regardless of its movement(Section 4.3), closing the window for further comparison. Additionally, the chassis library used in this pilot only supports sequenced heading/duration commands, so testing continuous curvature would require modification in the control layer to support continuous per-wheel velocity control, which is beyond this pilot cycle and can be included in future studies. The observed representation misalignment issues may stem from the methodological choice of this pilot to treat the subject as someone to test a hypothesis on, rather than as a participant who could actively shape the initial design of robot's behaviors, echoing Bobu et al.'s \cite{10.1145/3610977.3634987} critique of treating humans as static data sources instead of active collaborators in representation learning. Inspired by their work, future research could open an alternative direction, exploring real‑time predictive models of dog behavior to inform a robot’s representation on the fly. Ehsani et al \cite{8578524} has demonstrated that dog movement can be predicted directly from visual and inertial data.

More broadly, the pilot surfaces a structural challenge for AMI evaluation with companion animals: Proxy stimuli may not retain similar properties to be generalizable to the real-world referents. The brevity of coded behavioral durations in this pilot (typically under 5 seconds per behavior) stands in contrast to the sustained, cyclical retreat-and-freezing patterns anecdotally observed in the participant's real-world encounters with delivery robots, which have typically lasted for a minute or more. One possible explanation is that the evaluation was conducted in the participant's familiar residential home, which represents a low-stimulus environment relative to the urban sidewalk contexts where delivery robot encounters naturally occur. Given that dogs are highly context-sensitive animals, behavioral responses observed in a familiar home setting may not generalize to the arousal levels, spatial constraints, and competing stimuli present in real-world encounters. These asymmetric reactions to the proxy stimuli versus the real-world referents also suggest a gap that implicates acoustic profile, scale, and movement consistency as candidate dimensions along which the two diverge. Therefore, future work should also investigate which properties of real delivery robots sustain aversive responses in companion dogs.

\section{Limitations}
This pilot was only able to constitute a single completed cycle of an intended iterative research through design process with a single participant. The participant's rapid within-session habituation (Section 4) meant that only one curve-out design could be evaluated before her behavioral response to the maneuver changed, precluding within-study comparison across multiple trajectory variations. This suggests that for future RtD work with individual animal participants, habituation timelines may constrain how many design iterations a single participant can meaningfully evaluate within a session, and multi-iteration RtD in this context may require either multiple animal participants or longer inter-session gaps to mitigate habituation effects between design variants.

We also acknowledge the limitation of relying fully on real-time qualitative observation. This required simultaneously observing the participant and recording descriptive notes, and recorded behavior durations may therefore be systematically compressed relative to actual behavior length, as attention divided between observation and note-taking could delay or truncate the recorded offset of a behavior. Future studies should incorporate audio-visual recording to improve coding precision and support inter-rater reliability.

This pilot also suffered from the limitation of proxy validity in presenting generalizable findings. The proxy platform differs substantially from commercial delivery robots in scale, speed, and movement pattern: it is markedly smaller than deployed delivery robots and travels at approximately 0.67 mph, compared to the 3–5 mph walking speed reported for commercial delivery robots \cite{uber2026robots}. And the hardware only allows one to execute movement in a segmented fashion, substantially differing from commercial robots'smoothly steering movement pattern. These differences mean the pilot's proxy stimulus diverges meaningfully from the real-world referent it is intended to emulate. A future study should investigate whether using a more closely aligned proxy would close this gap.

\bibliographystyle{ACM-Reference-Format}
\bibliography{reference}

\end{document}